 \documentclass[pmlr,twocolumn]{jmlr} 

\usepackage{booktabs}
\usepackage[load-configurations=version-1]{siunitx} 
 
\usepackage{bm}

\theorembodyfont{\upshape}
\theoremheaderfont{\scshape}
\theorempostheader{:}
\theoremsep{\newline}

\jmlrvolume{ML4H Extended Abstract Arxiv Index}
\jmlryear{2020}
\jmlrsubmitted{2020}
\jmlrpublished{}
\jmlrworkshop{Machine Learning for Health (ML4H) 2020} 

\title[Decomposing Semantic Features of Medical Images]{Decomposing Normal and Abnormal Features of \titlebreak Medical Images for Content-based Image Retrieval}

\author{%
\Name{Kazuma Kobayashi} \Email{kazumkob@ncc.go.jp}\\
\addr National Cancer Center Research Institute, Japan\\
\Name{Ryuichiro Hataya} \Email{hataya@nlab.ci.i.u-tokyo.ac.jp}\\
\addr Graduate School of Information Science and Technology, The University of Tokyo, Japan\\
\Name{Yusuke Kurose} \Email{kurose@mi.t.u-tokyo.ac.jp}\\
\Name{Tatsuya Harada} \Email{harada@mi.t.u-tokyo.ac.jp}\\
\addr Research Center for Advanced Science and Technology, The University of Tokyo, Japan\\
\Name{Ryuji Hamamoto} \Email{rhamamot@ncc.go.jp}\\
\addr National Cancer Center Research Institute, Japan\\
}

\begin{document}

\maketitle

\begin{abstract}
Medical images can be decomposed into normal and abnormal features, which is considered as the {\it compositionality}. Based on this idea, we propose an encoder-decoder network to decompose a medical image into two discrete latent codes: a {\it normal anatomy code} and an {\it abnormal anatomy code}. Using these latent codes, we demonstrate a similarity retrieval by focusing on either normal or abnormal features of medical images. 
\end{abstract}

\section{Introduction}
\label{sec:intro}



In medical imaging, the characteristics purely derived from a disease should reflect the extent to which abnormal findings deviate from the normal features that would have existed. Indeed, physicians often need corresponding normal images without abnormal findings of interest or, conversely, images that contain similar abnormal findings regardless of normal anatomical context. This is called comparative diagnostic reading of medical images, which is essential for a correct diagnosis. To support the comparative diagnostic reading, content-based image retrieval (CBIR) utilizing either normal or abnormal features will be useful.


Here, we define this two-tiered nature of normal and abnormal features as the {\it compositionality} of medical images. Subsequently, we consider a method to decompose a medical image into two low-dimensional representations, where the two latent codes representing normal and abnormal anatomies should be collective for reconstructing the original image (Figure 1). 

To our best knowledge, few studies have focused on the compositionality of medical images. Recently, image-to-image translation techniques mainly derived from CycleGAN \citep{cyclegan8237506} were exploited to disentangle the domain-specific variations of medical images \citep{XIA2020101719, ADN8788607, vorontsov2019semisupervised}. For example, Tian et al. successfully transformed an input image with pathology into a normal-appearing image by leveraging several cycle-consistency losses \citep{XIA2020101719}. However, these approaches did not treat the compositionality in a separable manner, i.e., the latent spaces of these methods were not explicitly designed for down-stream tasks.

In this paper, we propose an encoder-decoder network to project a medical image into a pair of latent spaces, each of which produces a {\it normal anatomy code} and an {\it abnormal anatomy code}. Using these latent codes, our CBIR framework can retrieve images by focusing on either normal or abnormal features, providing excellent performances in qualitative evaluations.

\begin{figure}[t]
  \centering
  \label{fig:concept_of_segregation}
  \includegraphics[width=\linewidth]{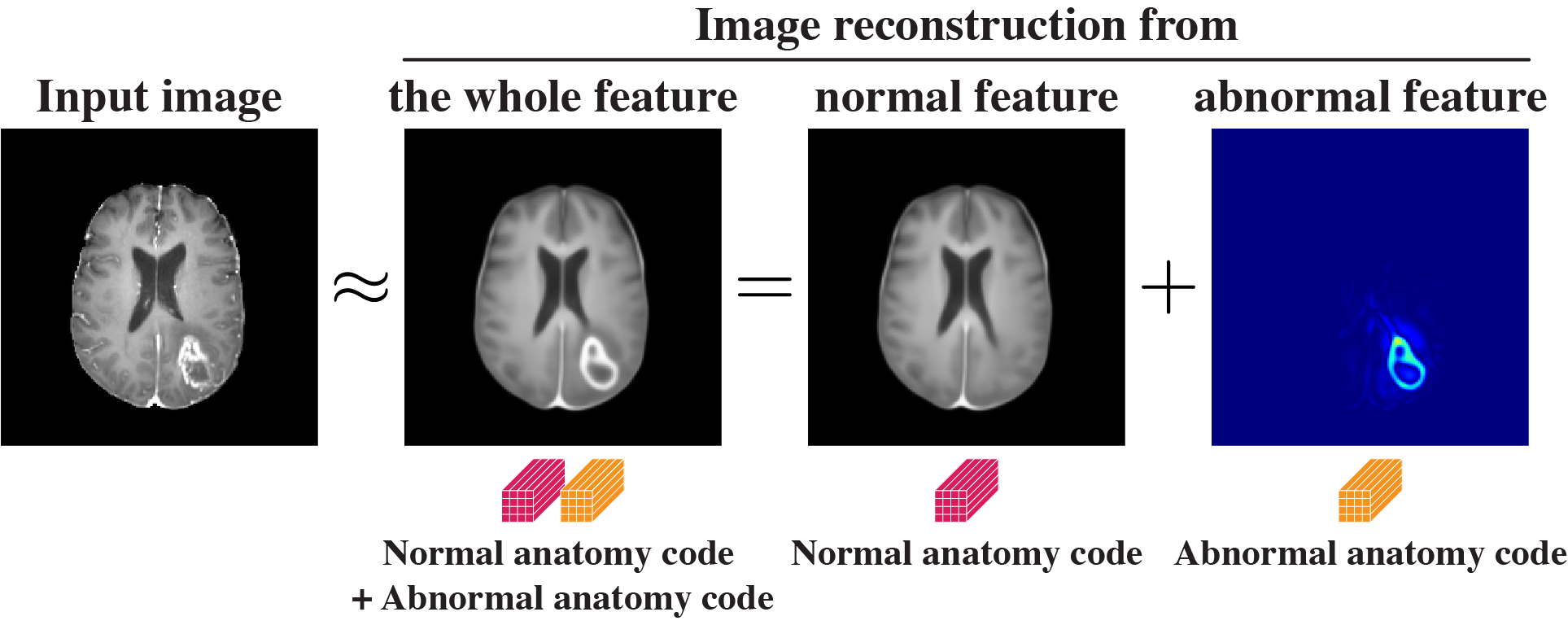}
  \caption{Our concept of the compositionality of medical images, indicating that the entire image should be composed of the representation of normal and abnormal features.}
\end{figure}


\begin{figure*}[t]
  \centering
  \label{fig:model_overview}
  \includegraphics[width=0.8\linewidth]{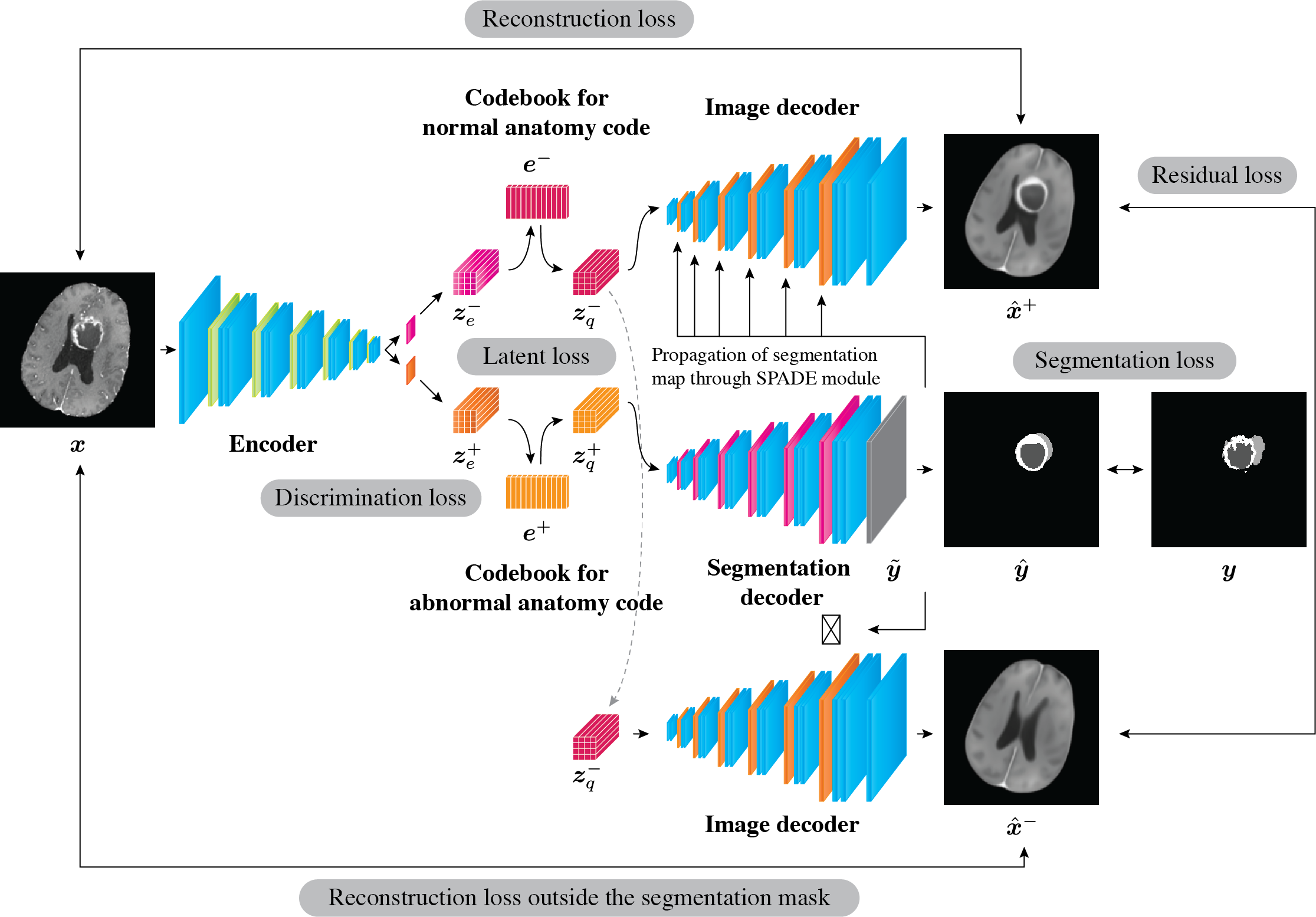}
  \caption{Schematic of the proposed encoder-decoder network.}
\end{figure*}


\begin{figure*}[t]
  \centering
  \label{fig:cbir_result}
  \includegraphics[width=0.8\linewidth]{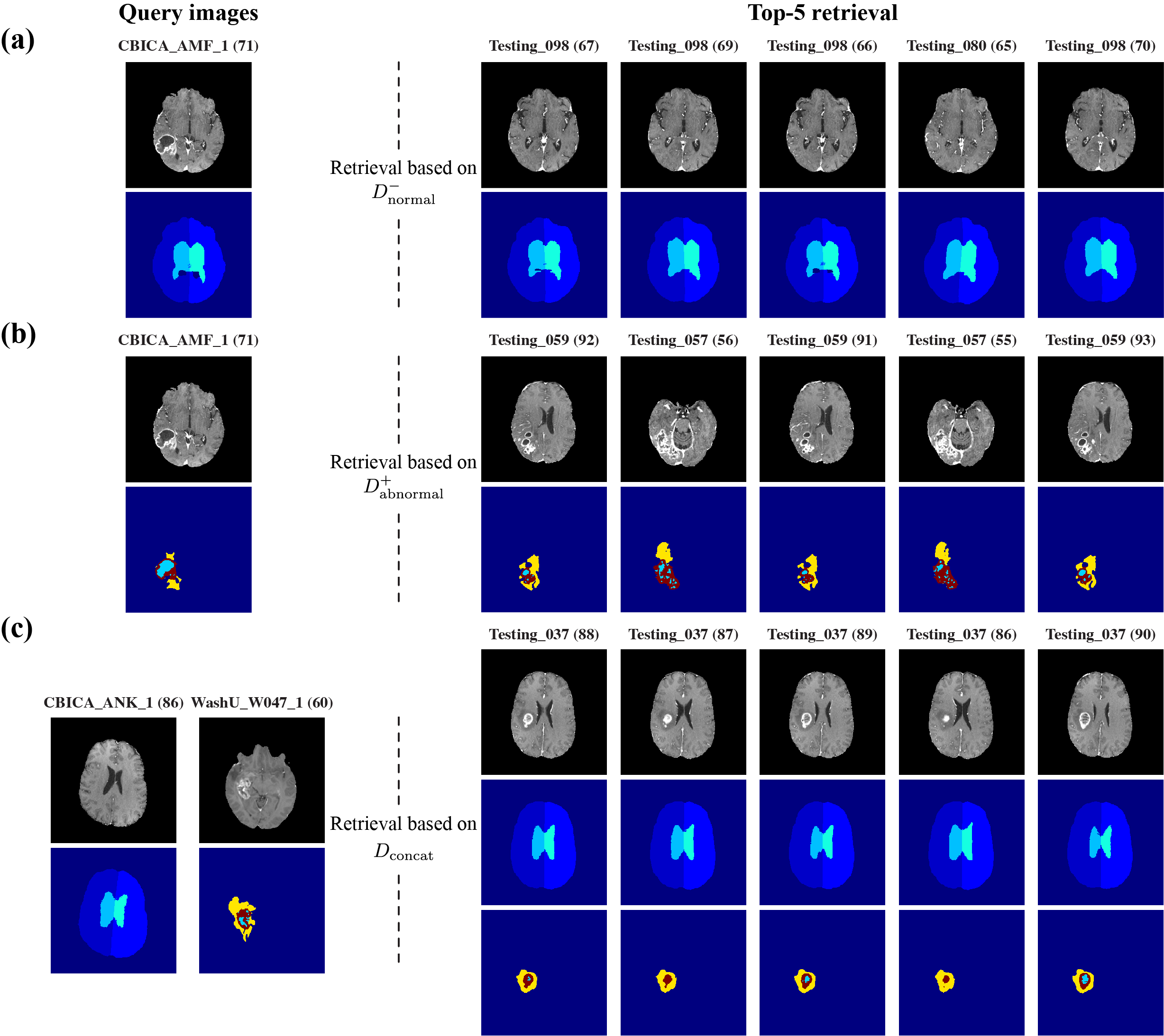}
  \caption{Example results of content-based image retrieval based on similarities of decomposed latent codes. Images are shown with normal or abnormal anatomical labels that are corresponded to the semantics utilized in the retrieval. Although the image retrieval employed latent codes instead of label information, the retrieved images accompanied similar labels to those of the query images. Patient identifiers with slice numbers are noted.}
\end{figure*}

\section{Methodology}
\label{sec:method}

The proposed network consists of an encoder and two decoders, a segmentation decoder and an image decoder (Figure 2). A pair of discrete latent spaces exists at the bottom that produces normal and abnormal anatomy codes, separately. See \appendixref{apd:detailed_architecture} for the details of the network architecture.

{\bf Feature encoding: }The encoder uses a two-dimensional medical image $\bm{x} \in \mathbb{R}^{C \times H \times W}$ as an input and maps it into two latent representations, $\bm{z}_e^- \in \mathbb{R}^{D \times H^{\prime} \times W^{\prime}}$ and $\bm{z}_e^+ \in \mathbb{R}^{D \times H^{\prime} \times W^{\prime}}$, where $\bm{z}_e^-$ and $\bm{z}_e^+$ correspond to the features of normal and abnormal anatomies, respectively. We used $\bm{z}^\mp_e$ to represent both features. Subsequently, vector quantization was used to discretize $\bm{z}^\mp_e$. Namely, each elemental vector $\bm{z}^\mp_{e_k} \in \mathbb{R}^{D}$ was replaced with the closest code vector in each codebook $\bm{e}^\mp \in \mathbb{R}^{D \times K}$ comprising $K$ code vectors. The codebooks were updated as VQ-VAEs \citep{oord2017neural, razavi2019generating}. We denote the quantized vector of $\bm{z}_e^\mp$ as $\bm{z}_q^\mp$. Here, $\bm{z}_q^-$ is referred to as the normal anatomy code, and $\bm{z}_q^+$ as the abnormal anatomy code.

{\bf Feature decoding: }The segmentation decoder uses abnormal anatomy code $\bm{z}_q^+$ as input and outputs segmentation label $\hat{\bm{y}} \in \mathbb{R}^{C^\prime \times H \times W}$. Meanwhile, the image decoder $f$ performs conditional image generation using the {\it spatially-adaptive normalization} (SPADE) \citep{park2019SPADE}. SPADE is designed to propagate semantic layouts to the process of synthesizing images (\appendixref{apd:spade}). The image decoder uses the normal anatomy code $\bm{z}_q^-$ as its primary input. When the image decoder is encouraged to reconstruct the entire input image $\hat{\bm{x}}^+$, the logit of the segmentation decoder $\tilde{\bm{y}}$ is transmitted to each layer of the image decoder via the SPADE modules ($f(\bm{z}_q^-, \tilde{\bm{y}}) = \hat{\bm{x}}^+$). When {\it null} information, where $\tilde{\bm{y}}$ is filled with 0s, is propagated to the SPADE modules, normal-appearing image $\hat{\bm{x}}^-$ is generated by the image decoder ($f(\bm{z}_q^-, \bm{0}) = \hat{\bm{x}}^-$).

{\bf Learning objectives: }We defined several loss functions (see \appendixref{apd:loss_functions} for the details): latent loss $L_\mathrm{lat}$ for optimizing the encoder and the codebooks, discrimination loss $L_\mathrm{dis}$ for the encoder to identify the presence of abnormality, segmentation loss $L_\mathrm{seg}$ for the segmentation decoder, reconstruction loss $L_\mathrm{rec}$, and residual loss $L_\mathrm{res}$ for the conditioned image generation performed in coordination between the two decoders. The overall objective can be summarized as follows: $L_\mathrm{total} = L_\mathrm{lat} + \lambda_1 L_\mathrm{dis} + \lambda_2 L_\mathrm{seg} + \lambda_3 L_\mathrm{rec} + \lambda_4 L_\mathrm{res}$, where $\lambda$s are used for balancing the terms. The details of learning configuration and an example of the model training result are presented in \appendixref{apd:training_setting} and \appendixref{apd:model_training_result}, respectively. 

{\bf Content-based image retrieval: }After the training, the encoder learned to decompose medical images into normal and abnormal anatomy codes. We defined three measurements according to the types of latent codes as follows: $D^{-}_\mathrm{normal}$ for normal anatomy codes, $D^{+}_\mathrm{abnormal}$ for abnormal anatomy codes, and $D_\mathrm{concat}$ for concatenated codes of the two codes. The L2 distance was calculated between the query and reference latent codes (see \appendixref{apd:cbir_overview} for the overview of the proposed CBIR method). 

\section{Dataset}

We used brain magnetic resonance (MR) images with gliomas from the 2019 BraTS Challenge \citep{6975210brats, Bakas2017, TCGAGBM, TCGALGG}, containing a training dataset with 355 patients, a validation dataset with 125 patients, and a test dataset with 167 patients. Among T1, gadolinium (Gd)-enhancing T1, T2, and FLAIR sequences, only the Gd-enhanced T1-weighted sequence was used. The training dataset contained three segmentation labels of abnormality: Gd-enhancing tumor (ET), peritumoral edema (ED), and necrotic and non-enhancing tumor core (TC). We used the training dataset to train the networks. Further, each image in the validation and test dataset was segmented into six normal anatomical labels (left and right cerebrum, cerebellum, and ventricles) and three abnormal labels (ET, ED, and TC). The validation and test datasets were used as query and reference datasets, respectively, for the performance evaluation of CBIR.

\section{Results}

Example CBIR results showing 5 images with the closest latent codes based on $D^{-}_\mathrm{normal}$,  $D^{+}_\mathrm{abnormal}$, and $D_\mathrm{concat}$ are presented in Figure 3. Distance calculation based on normal anatomy codes $D^{-}_\mathrm{normal}$ retrieved images with similar normal anatomical labels irrespective of gross abnormalities (Figure 3a). Distance calculation based on abnormal anatomy codes $D^{+}_\mathrm{abnormal}$ retrieved images with similar abnormal anatomical labels (Figure 3b). Note that the variety of normal anatomical contexts of the retrieved images. In the calculation using $D_\mathrm{concat}$, the query latent code was made from a combination of the normal anatomy code of the left image (CBICA\_ANK\_1) and the abnormal anatomy code of the right image (WashU\_W047\_1) (Figure 3c). Note that normal anatomies and abnormal anatomies of the retrieved images resemble those of the left query image and right query image, respectively. 


\section{Conclusion}

We demonstrated the CBIR algorithm focusing on the semantic composites of medical imaging. This application can be useful to support comparative diagnostic reading, which is essential for a correct diagnosis. We will further evaluate the quantitative performance of the proposed method. 

\clearpage

\acks{We are grateful to Dr. Ken Asada, Dr. Ryo Shimoyama, and Dr. Mototaka Miyake for helpful discussions. The authors thank the members of the Division of Molecular Modification and Cancer Biology of the National Cancer Center Research Institute for their kind support. The RIKEN AIP Deep Learning Environment (RAIDEN) supercomputer system was used in this study to perform computations. 

{\bf Funding:} This work was supported by JST CREST (Grant Number JPMJCR1689), JST AIP-PRISM (Grant Number JPMJCR18Y4), and JSPS Grant-in-Aid for Scientific Research on Innovative Areas (Grant Number JP18H04908).

{\bf Competing interests:} Kazuma Kobayashi and Ryuji Hamamoto have received research funding from Fujifilm Corporation.}

\bibliography{reference}

\appendix

\section{Detailed Network Architecture of the Proposed Model}\label{apd:detailed_architecture}

Table A.1 demonstrates the detailed architecture of the encoder. Table A.2 presents the shared architecture between the image and segmentation decoders. The two decoders share most of the components except for the normalization function, where the image decoder utilizes batch normalization \citep{batchnorm_icml15}, and the segmentation decoder exploits SPADE \citep{park2019SPADE}. 

\setcounter{table}{0}
\renewcommand\thetable{\thesection.\arabic{table}}
\setcounter{figure}{0}
\renewcommand\thefigure{\thesection.\arabic{figure}}

\begin{table}[htbp]
\label{tab:encoder_architecture}
\centering
\caption{The detailed architecture of the encoder.} 
\resizebox{\linewidth}{!}{\begin{tabular}{lcc}
\hline
\textbf{Module}                                                     & Activation                                                                                                                                                                                                         & Output shape                                                                                                                \\ \hline
\begin{tabular}[c]{@{}l@{}}Input image\\ Conv\\ AvgPool\end{tabular} & \begin{tabular}[c]{@{}c@{}}\\$\begin{bmatrix} 5 \times 5 & 32 \end{bmatrix}$\\ -\end{tabular}                                                                                                                        & \begin{tabular}[c]{@{}c@{}}$1 \times 256 \times 256$\\ $32 \times 256 \times 256$\\ $32 \times 128 \times 128$\end{tabular} \\ \hline
\begin{tabular}[c]{@{}l@{}}Res-block\\ \\ AvgPool\end{tabular}       & \begin{tabular}[c]{@{}c@{}}$\begin{bmatrix} 3 \times 3 & 64 \\ 3 \times 3 & 64 \end{bmatrix}$\\ -\end{tabular}                                                                                                     & \begin{tabular}[c]{@{}c@{}}$64 \times 128 \times 128$\\ \\ $64 \times 64 \times 64$\end{tabular}                            \\ \hline
\begin{tabular}[c]{@{}l@{}}Res-block\\ \\ AvgPool\end{tabular}       & \begin{tabular}[c]{@{}c@{}}$\begin{bmatrix} 3 \times 3 & 128 \\ 3 \times 3 & 128 \end{bmatrix}$\\ -\end{tabular}                                                                                                   & \begin{tabular}[c]{@{}c@{}}$128 \times 64 \times 64$\\ \\ $128 \times 32 \times 32$\end{tabular}                            \\ \hline
\begin{tabular}[c]{@{}l@{}}Res-block\\ \\ AvgPool\end{tabular}       & \begin{tabular}[c]{@{}c@{}}$\begin{bmatrix} 3 \times 3 & 256 \\ 3 \times 3 & 256 \end{bmatrix}$\\ -\end{tabular}                                                                                                   & \begin{tabular}[c]{@{}c@{}}$256 \times 32 \times 32$\\ \\ $256 \times 16 \times 16$\end{tabular}                            \\ \hline
\begin{tabular}[c]{@{}l@{}}Res-block\\ \\ AvgPool\end{tabular}       & \begin{tabular}[c]{@{}c@{}}$\begin{bmatrix} 3 \times 3 & 512 \\ 3 \times 3 & 512 \end{bmatrix}$\\ -\end{tabular}                                                                                                   & \begin{tabular}[c]{@{}c@{}}$512 \times 16 \times 16$\\ \\ $512 \times 8 \times 8$\end{tabular}                              \\ \hline
\begin{tabular}[c]{@{}l@{}}Split\end{tabular}                        & \begin{tabular}[c]{@{}c@{}}$\begin{bmatrix} 3 \times 3 & 64  \end{bmatrix}$, $\begin{bmatrix} 3 \times 3 & 64\end{bmatrix}$\end{tabular}                                                                         & \begin{tabular}[c]{@{}c@{}}$64 \times 8 \times 8$, $64 \times 8 \times 8$\end{tabular}          \\ \hline

\end{tabular}}
\end{table} 

\begin{table}[htbp]
\label{tab:decoder_architecture}
\centering
\caption{Detailed architecture shared between the image and segmentation decoders.} 
\resizebox{\linewidth}{!}{\begin{tabular}{lcc}
\hline
\textbf{Module}                                                & Activation                                                                                                                                                     & Output shape                                                                                       \\ \hline
Latent representation                                           & -                                                                                                                                                              & $64 \times 8 \times 8$                                                                            \\ \hline
Conv-block                                                      & $\begin{bmatrix} 3 \times 3 & 512 \\ 3 \times 3 & 512 \end{bmatrix}$                                                                                           & $512 \times 8 \times 8$                                                                            \\ \hline
\begin{tabular}[c]{@{}l@{}}Upsample\\ (SPADE-)Res-block \\ \\ \end{tabular} & \begin{tabular}[c]{@{}c@{}}$\begin{bmatrix} 3 \times 3 & 256\end{bmatrix}$\\ $\begin{bmatrix} 3 \times 3 & 256 \\ 3 \times 3 & 256 \end{bmatrix}$\end{tabular} & \begin{tabular}[c]{@{}c@{}}$256 \times 16 \times 16$\\ $256 \times 16 \times 16$ \\ \\ \end{tabular}   \\ \hline
\begin{tabular}[c]{@{}l@{}}Upsample\\ (SPADE-)Res-block \\ \\ \end{tabular}  & \begin{tabular}[c]{@{}c@{}}$\begin{bmatrix} 3 \times 3 & 128\end{bmatrix}$\\ $\begin{bmatrix} 3 \times 3 & 128 \\ 3 \times 3 & 128 \end{bmatrix}$\end{tabular} & \begin{tabular}[c]{@{}c@{}}$128 \times 32 \times 32$\\ $128 \times 32 \times 32$ \\ \\ \end{tabular}   \\ \hline
\begin{tabular}[c]{@{}l@{}}Upsample\\ (SPADE-)Res-block \\ \\ \end{tabular}  & \begin{tabular}[c]{@{}c@{}}$\begin{bmatrix} 3 \times 3 & 64\end{bmatrix}$\\ $\begin{bmatrix} 3 \times 3 & 64 \\ 3 \times 3 & 64 \end{bmatrix}$\end{tabular}    & \begin{tabular}[c]{@{}c@{}}$64 \times 64 \times 64$\\ $64 \times 64 \times 64$ \\ \\ \end{tabular}     \\ \hline
\begin{tabular}[c]{@{}l@{}}Upsample\\ (SPADE-)Res-block \\ \\ \end{tabular}  & \begin{tabular}[c]{@{}c@{}}$\begin{bmatrix} 3 \times 3 & 32\end{bmatrix}$\\ $\begin{bmatrix} 3 \times 3 & 32 \\ 3 \times 3 & 32 \end{bmatrix}$\end{tabular}    & \begin{tabular}[c]{@{}c@{}}$32 \times 128 \times 128$\\ $32 \times 128 \times 128$ \\ \\ \end{tabular} \\ \hline
\begin{tabular}[c]{@{}l@{}}Upsample\\ (SPADE-)Res-block \\ \\ \end{tabular}  & \begin{tabular}[c]{@{}c@{}}$\begin{bmatrix} 3 \times 3 & 16\end{bmatrix}$\\ $\begin{bmatrix} 3 \times 3 & 16 \\ 3 \times 3 & 16 \end{bmatrix}$\end{tabular}    & \begin{tabular}[c]{@{}c@{}}$16 \times 256 \times 256$\\ $16 \times 256 \times 256$ \\ \\ \end{tabular} \\ \hline
Conv                                                            & $\begin{bmatrix} 5 \times 5 & 1\end{bmatrix}$                                                                                                                  & $1 \times 256 \times 256$                                                                          \\ \hline
\end{tabular}}
\end{table} 

\section{SPADE Module for Propagation of Semantic Segmentation Map}\label{apd:spade}

The detailed architecture of the SPADE module is shown in Figure B.1.


\begin{figure*}[t]
  \centering
  \label{fig:spade}
  \includegraphics[width=0.5\linewidth]{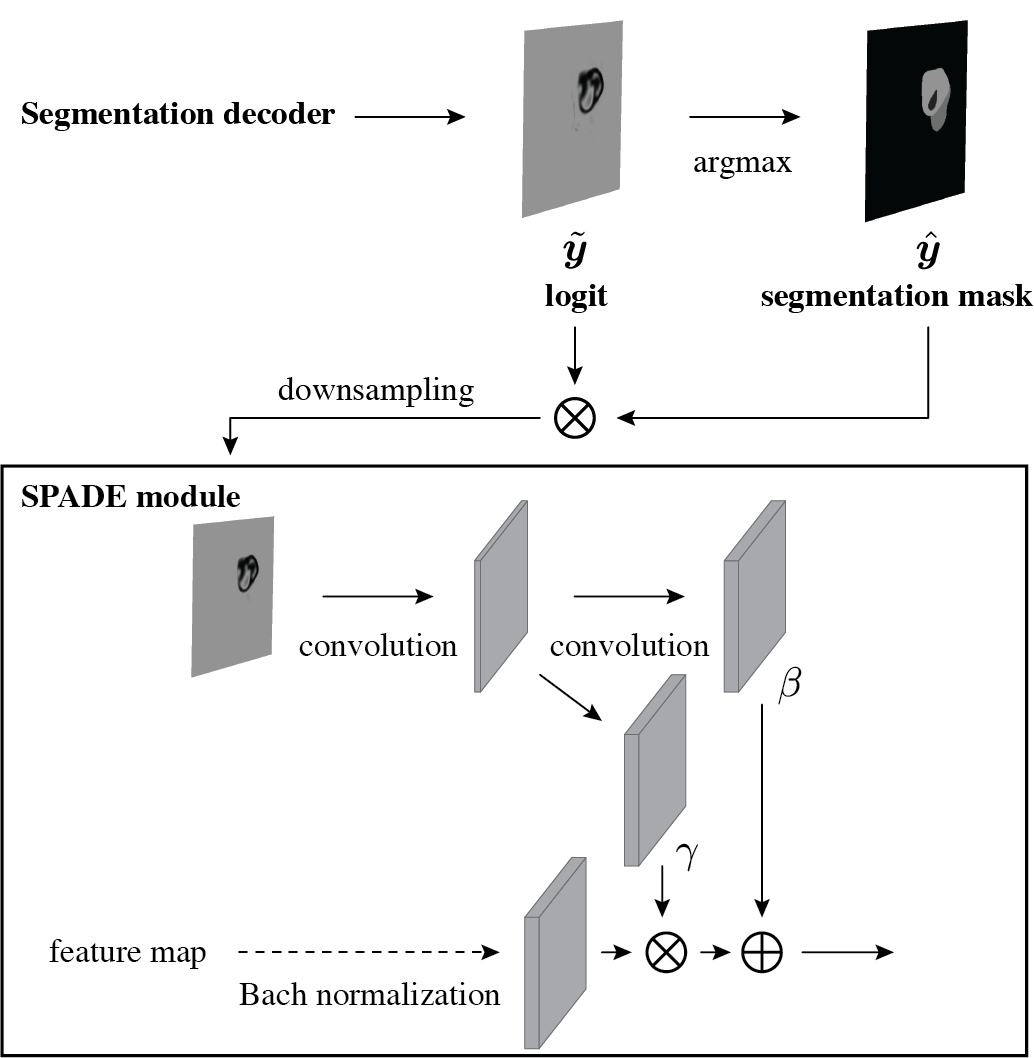}
  \caption{Logit $\tilde{\bm{y}}$ applied by the segmentation mask $\hat{\bm{y}}$ is further downsampled to achieve resolutions corresponding to those of each layer in the image decoder. SPADE module propagates the semantic layout of abnormalities into the image generation process.}
\end{figure*}

\section{Details of Learning Objectives}\label{apd:loss_functions}

Several loss functions were designed for the training. Hereinafter, we denote by $\mp$ to indicate that a particular term is either for the path based on the normal anatomy code ($-$) or abnormal anatomy code ($+$). 


{\bf Latent loss: }In the learning framework of the VQ-VAE \citep{oord2017neural, razavi2019generating}, the latent loss $L_\mathrm{lat}$ is optimized for acquiring latent embeddings for data samples. We define $L_\mathrm{lat}$ as a sum of $L^-_\mathrm{lat}$ and $L^+_\mathrm{lat}$ for the normal and abnormal anatomy codes, respectively, as follows:
\begin{equation} 
\begin{split}
L^\mp_\mathrm{lat} &= \|\mathrm{sg}[\bm{z}^\mp_e(x)] - \bm{e}^\mp\|^2_2 \\
                   &+ \beta \|\bm{z}^\mp_e(x) - \mathrm{sg}[\bm{e}^\mp]\|^2_2,
\end{split}
\end{equation} 
\begin{equation} 
L_\mathrm{lat} = L^{-}_\mathrm{lat} + L^{+}_\mathrm{lat},
\end{equation} 
where $\mathrm{sg}$ represents the stop-gradient operator that serves as an identity function at the forward computation time and has zero partial derivatives. During the training, the {\it codebook loss}, which is the first term in the equation above, updates the codebook variables by transferring the selected latent codes to the output of the encoder. Additionally, the {\it commitment loss}, which is the second term, encourages the output of the encoder to move closer to specific latent codes. 


{\bf Discrimination loss: }Because the input images do not always convey abnormal findings, the encoder must be able to distinguish the abnormalities. To implement a discriminative function in the encoder, we extend the commitment loss particularly for the abnormal anatomy code. The encoder is trained to minimize the commitment loss when abnormalities exist. Meanwhile, for normal input images, the encoder is encouraged to increase the term up to a threshold value of $\pi$. We define this loss function as the {\it discrimination loss} for the path to the abnormal anatomy code as follows: 
\begin{equation} 
L_\mathrm{dis} = \max(\pi - \|\bm{z}^+_e(x) - \mathrm{sg}[\bm{e}^+]\|^2_2, 0),
\end{equation} 
where $\pi$ is a positive scalar of the threshold.


{\bf Segmentation loss: }The segmentation decoder infers the segmentation labels, which are classified as $\mathcal{K} (= 3)$ abnormal segmentation categories in the training dataset. The loss function for the output of the segmentation decoder is a composite of the generalized Dice \citep{Sudre_2017} and cross-entropy losses as follows:
\begin{equation} 
L_\mathrm{dice} = 1 - 2 \frac{\sum_{k \in \mathcal{K}} w_k |\hat{\bm{y}}_{k} \cap \bm{y}_{k}|}{\sum_{k \in \mathcal{K}} w_k (|\hat{\bm{y}}_{k}| + |\bm{y}_{k}|)},
\end{equation} 
\begin{equation} 
L_\mathrm{entropy} = - \frac{1}{N} \sum_{k \in \mathcal{K}} \bm{y}_k \log \tilde{\bm{y}},
\end{equation} 
\begin{equation} 
L_\mathrm{seg} = L_\mathrm{dice} + L_\mathrm{entropy},
\end{equation} 
where $\tilde{\bm{y}}$ indicates the logit output of the segmentation decoder, $N$ is the number of pixels, and $w_k$ is determined as $w_k = \frac{1}{(\sum_N \bm{y}_k)^2}$ to mitigate the class imbalance problem.



{\bf Reconstruction loss: }To guarantee a difference between two types of generated images, $\hat{\bm{x}}^-$ and $\hat{\bm{x}}^+$, we applied a pixel-wise reconstruction loss based on the region of abnormality. Suppose $\bm{M} \in \{0, 1\}^{C \times H \times W}$ defines the mask, indicating that pixels with any abnormality labels are set to 1 and 0 otherwise, and $\overline{\bm{M}}$ is the complementary set of $\bm{M}$. Briefly, $\bm{M}$ presents the region of abnormality, and $\overline{\bm{M}}$ indicates the region of normal anatomy. Using these masks, the reconstruction loss $L_\mathrm{rec}$ is defined as follows: \begin{equation} 
\begin{split}
L^-_\mathrm{rec} &= \|\overline{\bm{M}} \odot \bm{x}^- - \overline{\bm{M}} \odot \bm{x}\|^2_2 \\
                            &+ (1 - \mathrm{SSIM}(\overline{\bm{M}} \odot \bm{x}^-, \overline{\bm{M}} \odot \bm{x})),
\end{split}
\end{equation} 
\begin{equation}
\begin{split}
L^+_\mathrm{rec} &= \|\bm{x}^+ - \bm{x}\|^2_2 \\
                            &+ (1 - \mathrm{SSIM}(\bm{x}^+, \bm{x})) \\
                            &+ \|{\bm{M}} \odot \bm{x}^+ - {\bm{M}} \odot \bm{x}\|^2_2 \\
                            &+ (1 - \mathrm{SSIM}({\bm{M}} \odot \bm{x}^+, {\bm{M}} \odot \bm{x})),
\end{split}
\end{equation} 
\begin{equation} 
L_\mathrm{rec} = L^-_\mathrm{rec} + L^+_\mathrm{rec},
\end{equation} 
where SSIM indicates the structural similarity \citep{Wang2004}, which is added to the L2 loss as a constraint.


{\bf Residual loss: }The image outside the region of abnormality must be the same between the two types of images, $\hat{\bm{x}}^-$ and $\hat{\bm{x}}^+$, generated by the image decoder to preserve the identity between corresponding regions. Therefore, we added a loss function to guarantee the similarity between $\hat{\bm{x}}^-$ and $\hat{\bm{x}}^+$ based on the normal regions,  indicated by $\overline{\bm{M}}$ as follows: 
\begin{equation} 
L_\mathrm{res} = \|\overline{\bm{M}} \odot \bm{x}^- - \overline{\bm{M}} \odot \bm{x}^+\|_1.
\end{equation} 

\section{Training Setting}\label{apd:training_setting}

All neural networks were implemented using Python 3.7 with PyTorch library 1.2.0 \citep{NEURIPS2019_9015} on an NVIDIA Tesla V100 graphics processing unit with CUDA 10.0. He initialization \citep{he2015delving} was applied to both the encoder and the decoder. Adam optimization \citep{kingma2014adam} was used with learning rates of $5 \times 10^{-3}$. Other hyperparameters were empirically determined as follows: batch size = 240, maximum number of epochs = 300, $\lambda_1 = 1.0$, $\lambda_2 = 1.0$, $\lambda_3 = 1.0 \times 10^4$, $\lambda_4 = 1.0$, $\lambda_5 = 1.0$, and $\pi = 10.0$. The input images were grayscale two-dimensional images with the size of $1 \times 256 \times 256$. The size of latent codebook was $512 \times 64$ ($= K \times D$). During training, data augmentation included horizontal flipping, random scaling, and rotation. 

\section{Training Results}\label{apd:model_training_result}
\setcounter{figure}{0}

The results of the model training at epoch 300 are shown in Figure E.1.


\begin{figure*}[t]
  \centering
  \label{fig:model_training_result}
  \includegraphics[width=\linewidth]{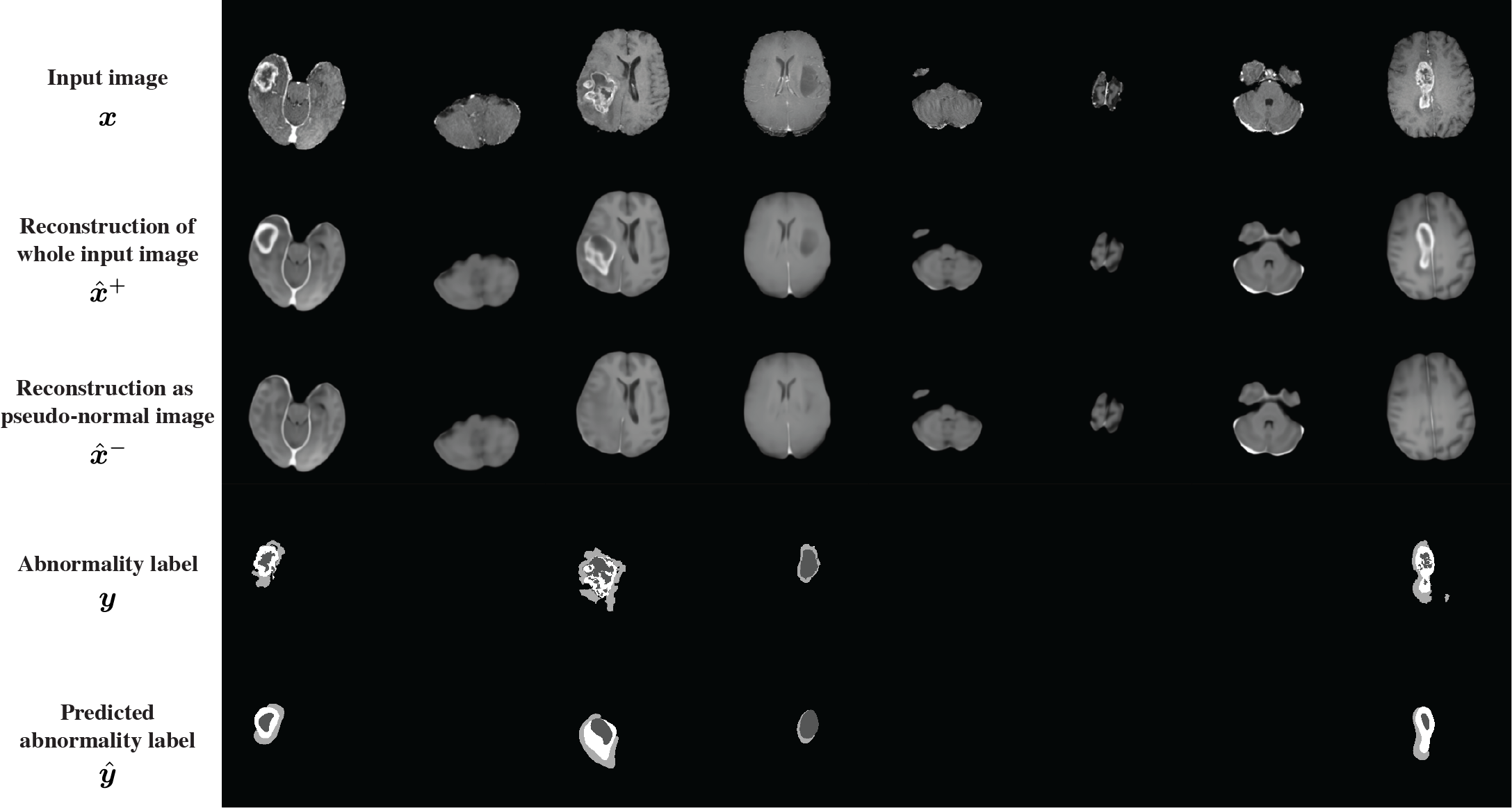}
  \caption{Results of model training. Entire input images $\hat{\bm{x}}^+$ (second row) were reconstructed based on both normal and abnormal anatomy codes, whereas reconstruction as pseudo-normal images $\hat{\bm{x}}^-$ (third row) were only on normal anatomy code. A clear distinction can be observed between $\hat{\bm{x}}^+$ and $\hat{\bm{x}}^-$ at abnormal regions, which existed in both $\bm{x}$ and $\hat{\bm{x}}^+$ but not in $\hat{\bm{x}}^-$. The fourth and fifth rows indicate ground-truth segmentation label $\bm{y}$ for abnormality (ET, TC, and ED) and prediction for labels $\hat{\bm{y}}$, respectively. The output of segmentation labels tended to be spherical and did not recover the detailed shape of each region. We assume this as a natural consequence since the compressed representation in the latent codes, which is advantageous for the computational cost of similarity search, did not have sufficient capacity to preserve the detailed feature in the input image as a trade-off.}
\end{figure*}

\section{Overview of Design of Content-based Image Retrieval}\label{apd:cbir_overview}
\setcounter{figure}{0}

An overview of the design of CBIR is presented in Figure F.1.


\begin{figure*}[t]
  \centering
  \label{fig:cbir_overview}
  \includegraphics[width=\linewidth]{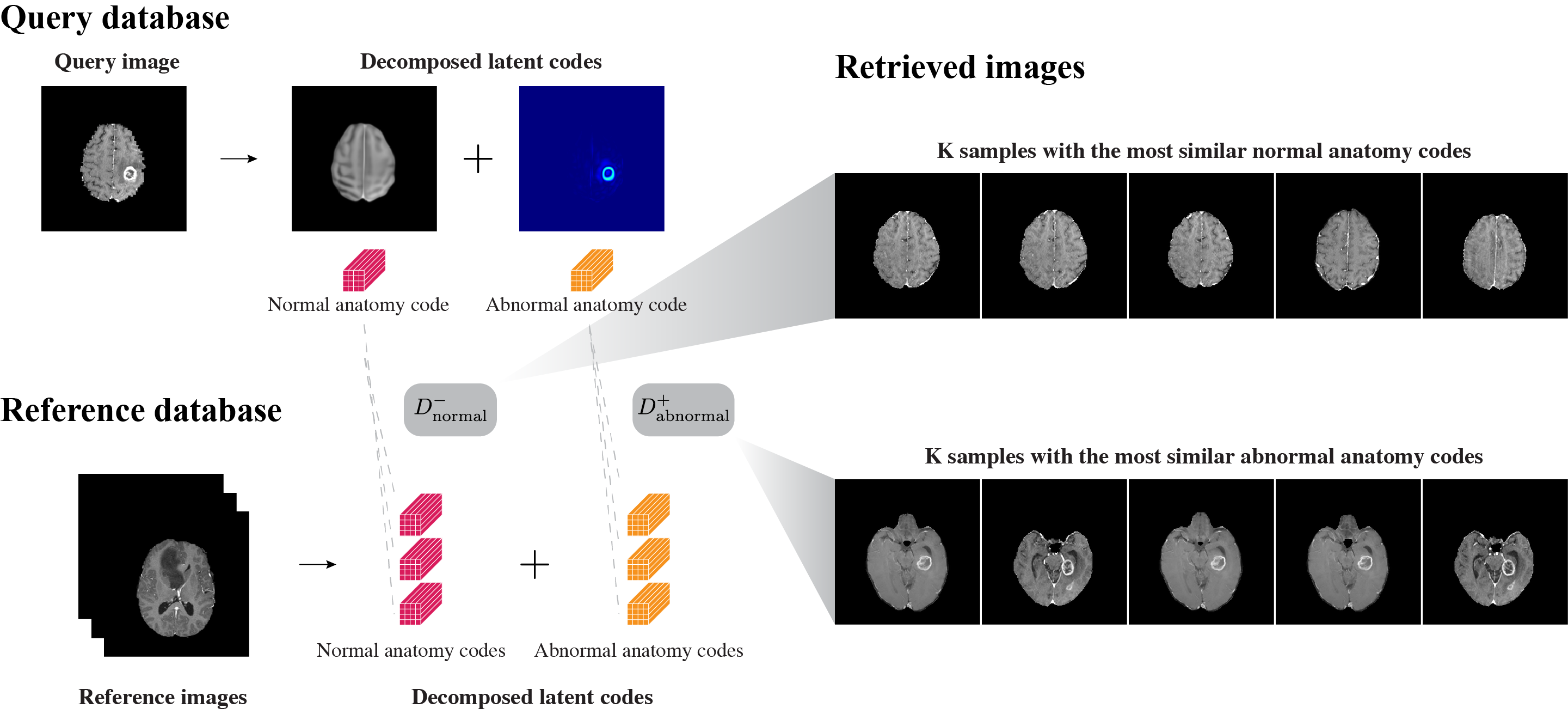}
  \caption{Content-based image retrieval was constructed on a per-image basis, i.e., each magnetic resonance volume was separated into slices along the axial axis. From the 2019 BraTS Challenge, the validation and test datasets were used as query and reference datasets, respectively. Every image in the reference dataset was decomposed into normal and abnormal anatomy codes in advance. Furthermore, a query image was decomposed into two latent codes. Subsequently, several reference images with the most similar latent codes were extracted from the reference database.}
\end{figure*}

\end{document}